# Non-Equilibrium Thermodynamic Extremal Principles During Filament Formation in ECM Memristors

*Justin Brutger and Xiao Shen*[*]

Department of Physics and Materials Science, University of Memphis, Memphis, TN, 38152

**ABSTRACT:**

Electrochemical metallization (ECM) memristors have potential applications in future neuromorphic computing hardware. The set, reset, and variable-resistance features of these devices originate in the formation and breakup of metal filaments in a solid-state electrolyte. While the performance characteristics of these devices are widely investigated, the driving principles behind the morphology of the filament formation process remain unclear. In this study, we propose an approach motivated by the extremal principles found in non-equilibrium thermodynamics and observe an entropy production and energy dissipation rate minimization during the filament-forming process in kinetic Monte Carlo simulations.

## I. INTRODUCTION

Resistive switching is an emerging area of study with wide-reaching applications in high-density memory storage and neuromorphic computing. Numerous types of resistive switching devices have been developed, with most designs relying on the formation and dissolution of conductive filaments as their mechanism for resistive switching.[1,2] Each design has its own benefits and drawbacks in terms of scalability, stability, and switching speed. A widespread adoption of resistive switching technology requires a device design that meets all three requirements. Although the device characteristics of resistive switching have been described well in the theoretical models of memristors,[3-5] the current theoretical understanding of the stability of these devices is inadequate.[6] Of particular need for such theoretical development in switching stability is the electrochemical metallization (ECM) memristor. The mechanism of filament formation in ECM memristors is through the electrically driven migration of diffused ions in a host matrix.[7,8] This migration is a stochastic process, and the ion drift that occurs during filament formation is the primary cause of the device's switching instability.[9] This switching instability is a major hurdle in ECM memristor design.

[*] xshen1@memphis.edu

A number of experimental techniques have been developed to remedy ECM memristor instability, such as nanorod insertion[10] or doping of the diffusive region.[11] However, the underlying theory of the factors driving ECM memristor stability is underdeveloped.[5,6] Previous research has linked filament stability and lifetime to the filament's morphology, and the device properties that produce different morphologies have been identified.[12,13] These filament morphologies are formed from a self-relaxation behavior present in a held filament,[14,15] and so represent the steady state behavior of the system. We propose that an alternative approach can be found in non-equilibrium thermodynamics, in which many extremal principles have been put forth to explain steady states in non-equilibrium systems. These extremal principles are varied, characterize steady states in terms of entropy production and energy dissipation, and typically rely on restrictive assumptions.[16] Despite their restrictiveness, extremal principles have found success in explaining the steady-state behavior of numerous systems, including resistive switching, heat and fluid flow, and negative-resistance solids.[17-19] Based on these approaches to non-equilibrium systems, we hypothesize that the filament morphology and stability in ECM memristors can be described by extremal principles and specifically examine the entropy production and energy dissipation behavior during filament formation and dissolution.

The application of extremal principles to filament-forming electrical devices has been explored previously for negative-resistance devices.[19] In more recent times, a minimum entropy production principle has been used to calculate the steady states of resistive switching in transition metal oxides.[17] These works assume the validity of an extremal principle and work towards deriving a steady state. The conditions for these principles to be valid were not explored. Since these extremal principles are particular to the conditions of the system involved, it would be instructive to start with a physically motivated steady state and determine what extremal principles are observed. Such an approach requires close attention in modeling all of the major processes in the system, and although filament formation and dissolution provide a simple picture of resistive switching in ECM devices, the physical phenomena that drive filament behavior are numerous, including thermal crystallization, charge trapping, and ionic migration.[20] A wide array of models have been developed for memristor simulation.[6,21,22] In this work, we follow the kinetic Monte Carlo approach used by Dirkmann et al. to model the filament formation and dissolution process,[23] with the additional calculation of the entropy production and energy dissipation rates. With these methods, we examine what extremal principles the ECM memristor obeys during the filament formation and dissolution processes.

## II. METHODS

In this kinetic Monte Carlo model, the atoms of the active electrode are arranged on a square grid and can undergo a selection of provided physical processes. At each simulation step, one of the atoms and processes is chosen to be carried out with a probability weighted by the reaction rate of that process. The advancement in simulation time is then determined by the chosen reaction rate as given as

$$\Delta t = -ln(r)/Q \, , r \in [0,1], \tag{1}$$

where r is a random number and Q is the sum of all possible reaction rates. These reaction rates are given by an Arrhenius law and are dependent on both the particular atom position and the overall configuration of the simulation at that moment. Following the approach of Dirkmann et al.,[21,23] the available physical processes for the atoms of the active electrode are depicted in Figure 1.

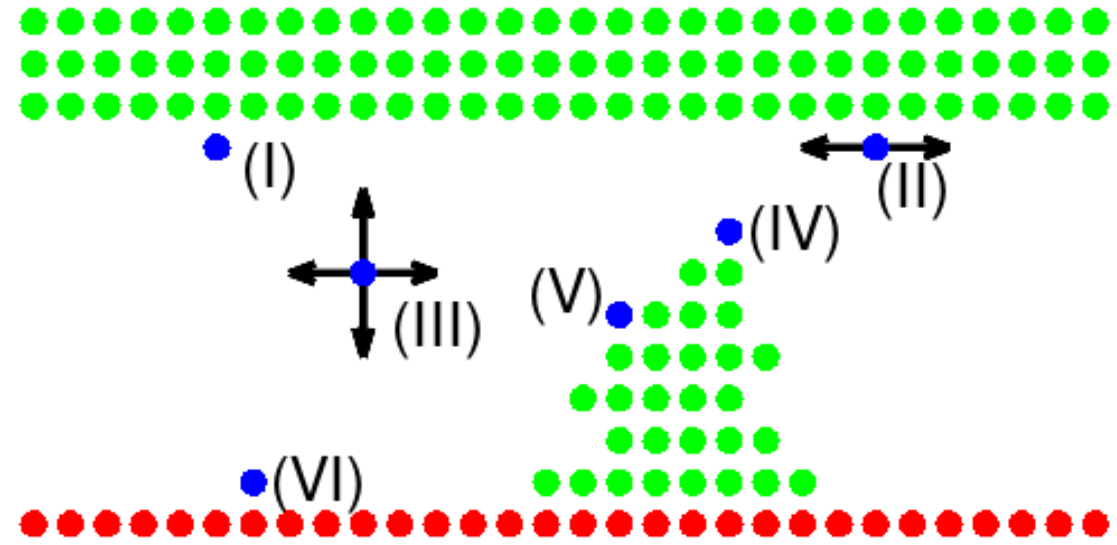

Fig. 1. Processes included within the simulation: (I) oxidation, (II) surface diffusion, (III) diffusion within host matrix, (IV) reduction at adatom site, (V) reduction at kink/hole site, and (VI) nucleation. Green: metal atoms on the active electrode and filament; blue: mobile ions in electrolyte; red: passive electrode.

The reaction rates are then given as

$$k = v_0 exp[-(E_a + ze\Delta\phi)/k_B T], \quad (2)$$

where $v_0$ is the phonon frequency, $k_B$ is the Boltzmann constant, $T$ is the temperature at the atom site, $e$ is the elementary charge, $z$ is the ionic charge number, $E_a$ is the inherent activation energy of the process, and $\Delta\phi$ is the change in the external electric potential between the migration sites. The electric and temperature properties are described using a continuum approach. The current is presumed to be ohmic and quasi-stationary, leading to the continuity equation

$$\nabla \cdot [\sigma(\vec{r})\nabla\phi(\vec{r})] = 0, \quad (3)$$

where $\sigma$ is the electrical conductivity and $\phi$ is the external electric potential. The temperature is calculated using the heat transfer equation as

$$c_p(\vec{r})\rho(\vec{r})\frac{\partial T(\vec{r},t)}{\partial t} - \nabla[\lambda(\vec{r})\nabla T(\vec{r},t)] = j^2(\vec{r},t)/\sigma(\vec{r}), \quad (4)$$

where $c_p$ is the specific heat, $\rho$ is the mass density, $T$ is the temperature, $\lambda$ is the thermal conductivity, $\boldsymbol{j}$ is the current density, and $\sigma$ is the electrical conductivity. Both Eq. 3 and Eq. 4 were solved by using the Gauss-Seidal and finite difference methods. The recurrence equation for Eq. 3 used a first-order central difference for the spatial derivative with periodic boundary conditions on the sides and Dirichlet boundary conditions on the top and bottom. The recurrence equation for Eq. 4 used a first-order backwards difference for the time derivative and a first-order central difference for the spatial derivative with the same boundary conditions as in Eq. 3. In addition, the entropy production rate of the system is calculated at each KMC step using the linear phenomenological approximation. The thermodynamic effects of heat conduction and Joule heating are considered, while the thermoelectric effects are assumed to be negligible. The entropy production rate per unit area is given in Eq. 5 for heat conduction and in Eq. 6 for Joule heating

$$\sigma_T = (\kappa\nabla^2 T)/T^2 \quad (5)$$

$$\sigma_{JH} = \rho j^2/T, \quad (6)$$

where $\kappa$ is the coefficient of heat conductivity, $T$ is the temperature, $\rho$ is the resistivity, and $\boldsymbol{j}$ is the current density.

The entropy production and energy dissipation behaviors are monitored during both the initial filament formation phase and the relaxation phase in an ECM memristor. To form a steady-state filament, the memristor is held at a constant current until the device resistance equilibrates. The initial simulation configuration consists of an 80 by 24 square lattice with a 10-atom-thick active electrode. A 5 by 3 atom prepatterning is placed on the inactive electrode as a seed to skip the nucleation process and to focus the growth on a single filament for analysis. The simulation parameters are shown in Table 1.

Table 1. Simulation parameters.

| **Physical Quantity** | **Value(s)** |
|---|---|
| Applied current | 3223 A |
| Boundary Temperature | 300 K |
| Conductivity $\sigma$ of host/electrode[21] | 100 /$6.3 \times 10^7$S/m |
| Ionic charge number $z$ | 1 |
| $E_{oxidation}$ of electrode material[21] | 0.65 eV |
| $E_{reduction}$ (surface[21]/kink/hole) | 0.62 / 0.60 / 0.58 eV |
| $E_{nucleation}$[21] | 0.81 eV |
| $E_{diffusion}$ through host matrix[21] | 0.61 eV |
| $E_{diffusion}$ along electrode surface[21] | 0.59 eV |
| Mass density $\rho$ of host[24]/electrode[25] | 4230 / 10490 W/(m K) |
| Heat capacity $c_p$ of host[26]/electrode[27] | 700 / 235 J/(kg K) |
| Thermal conductivity $\lambda$ of host/electrode | 3.5 / 214.5 kg/m$^3$ |

## III. RESULTS AND DISCUSSION

The results of the entropy production and energy dissipation rates averaged over a sample of 20 runs are shown in Figures 2(a) and 2(b). Two phases of behavior can be seen. The first is an initial decline in both rates, followed by a sudden transition to the second phase where both rates fluctuate around their steady state values. The first phase of behavior represents the minimization process as the system approaches the steady state entropy production and energy dissipation rates. The second phase of behavior represents the steady state of the system, wherein the entropy production and energy dissipation rates fluctuate diminishingly over time. Figure 2(c) shows the breakup of the contributions in the entropy production. Throughout the simulation, the entropy production due to Joule heating is comparable to the entropy produced by heat conduction, although the Joule heating entropy production is minimized at a slower rate.

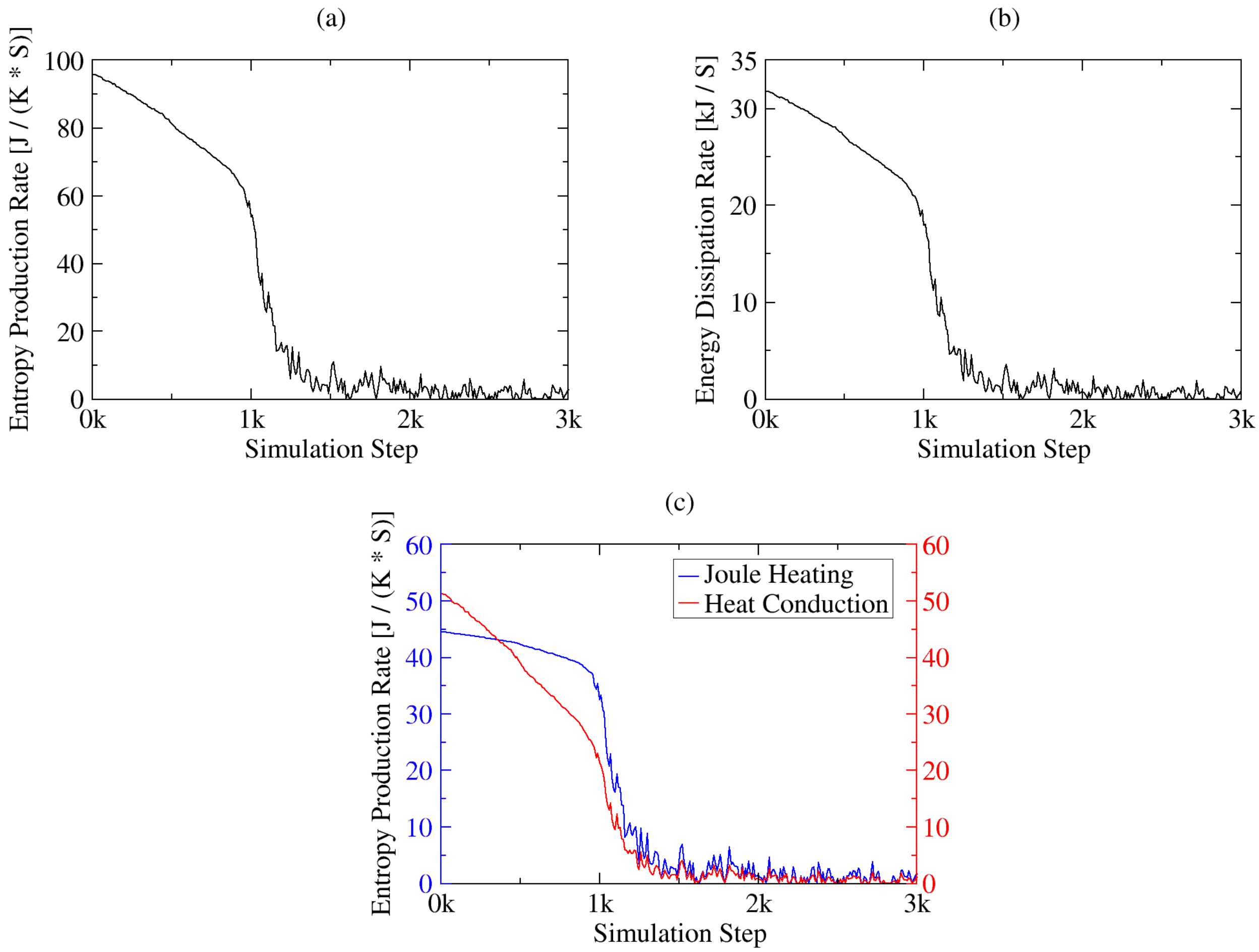

Fig. 2. Run-averaged extremal behavior. (a) Average entropy production rate. (b) Average energy dissipation rate. (c) Entropy production rate by source, averaged over simulation runs.

The results of the filament morphology and device resistance are shown in Figure 3. At the beginning, the constant current and high device resistance lead to a large voltage being applied, and the atoms on the active electrode rapidly oxidize and migrate to the inactive electrode due to the internal electric field, forming the initial filament connection. This corresponds to the first phase of behavior observed in the entropy production and energy dissipation rates. After the filament connects to the active electrode, the device resistance and applied voltage drop, leading the filament growth to become undirected, and the average filament morphology becomes stable. This corresponds to the second phase of behavior observed in the entropy production and energy dissipation rates.

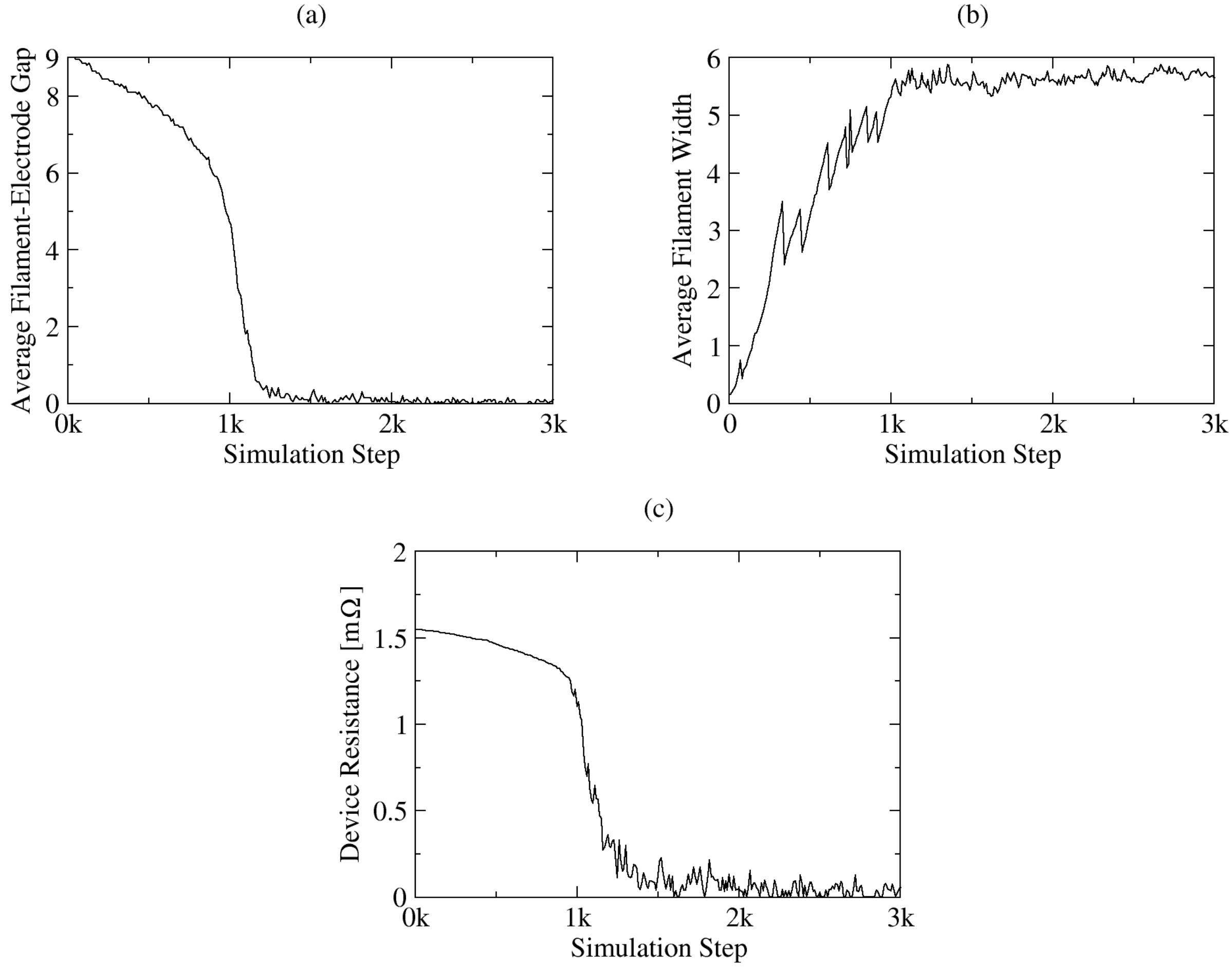

Fig. 3. Filament growth behavior averaged over all runs. (a) Average distance from the filament to the active electrode, measured in units of lattice sites. (b) Average filament width in terms of number of atoms. Filament width counting does not include the seed prepatterning or the side-branches that carry negligible current. (c) Device resistance.

The first phase of behavior corresponds to the initial filament formation. It is at this time that the filament bridges the distance between the two electrodes and also where the majority of the minimization occurs for the entropy production and energy dissipation rates. Initially, the low conductivity host matrix serves as a significant source of Joule heating that decreases slowly due to the initial slow filament growth rate. Since the current will need to pass through the low-conductivity material without a direct filament connection, the Joule heating remains high until the bridging is complete. In contrast, the entropy production from heat conduction is decreased steadily by the partial formation of the filament. The high thermal conductivity of the filament can effectively dissipate the Joule heating generated in the host matrix and decrease the temperature gradient without requiring a complete bridging. After this initial gradual filament growth, the electric field becomes concentrated enough to direct ionic diffusion towards the partial filament. This effect becomes stronger as the filament gap gets smaller. This quickly leads to the complete bridging of the filament. The sharp decrease of the entropy production at this point is directly due to the switch to the low resistance state that occurs as a consequence of the filament growth. Similarly, the decrease in energy dissipation rates is also due to the development of the low-resistance state.

The second phase of behavior corresponds to the fully connected filament. It is at this point that the filament is in its low-resistance state. In the ohmic model used in the simulation, this directly leads to a lower voltage difference between electrodes and a smaller electric field. Without a strong electric field, the filament morphology will undergo undirected growth and dissolution, which can be seen in the steady filament morphology parameters in the averaged run graphs in Figure 3 during this phase. This leads to a steady value of the average device resistance, and subsequently, the entropy production and energy dissipation rates will stop their monotonic descent in the first phase. The complete bridging is responsible for the final minimization of the entropy production and energy dissipation rates. With the bridged filament, the current will be confined to the high-conductivity pathway with minimal Joule heating occurring. Likewise, the entropy production from the heat conduction will be minimized through the vanishing of the thermal gradient in the high-conductivity filament.

While the steady-state behavior in the second phase is clear in the run-averaged results, individual simulation runs still exhibit a degree of instability in the system. The entropy production and energy dissipation rates for a selected individual run are shown in Figures 4(a) and 4(b). Here, it can be seen that the second phase of behavior is interspersed with alternative modes where the entropy production and energy dissipation rates momentarily increase. However, the timing is stochastic due to the undirected nature of the filament behavior during the second phase of behavior. This stochastic timing is reflected in the small fluctuations in the entropy production and energy dissipation behavior in the run-averaged data in Figures 2(a) and 2(b). The breakdown of entropy production rates by source during these alternate modes is also shown in Figure 4(c). During the momentary spikes in entropy production, the Joule heating and heat conduction terms behave as in the first phase of behavior. The behavior of entropy production outside these spikes is shown in Figure 4(d). It is clear that the Joule heating term dominates while the heat conduction term vanishes. The entropy production and energy dissipation rates continue to be minimized in this phase diminishingly.

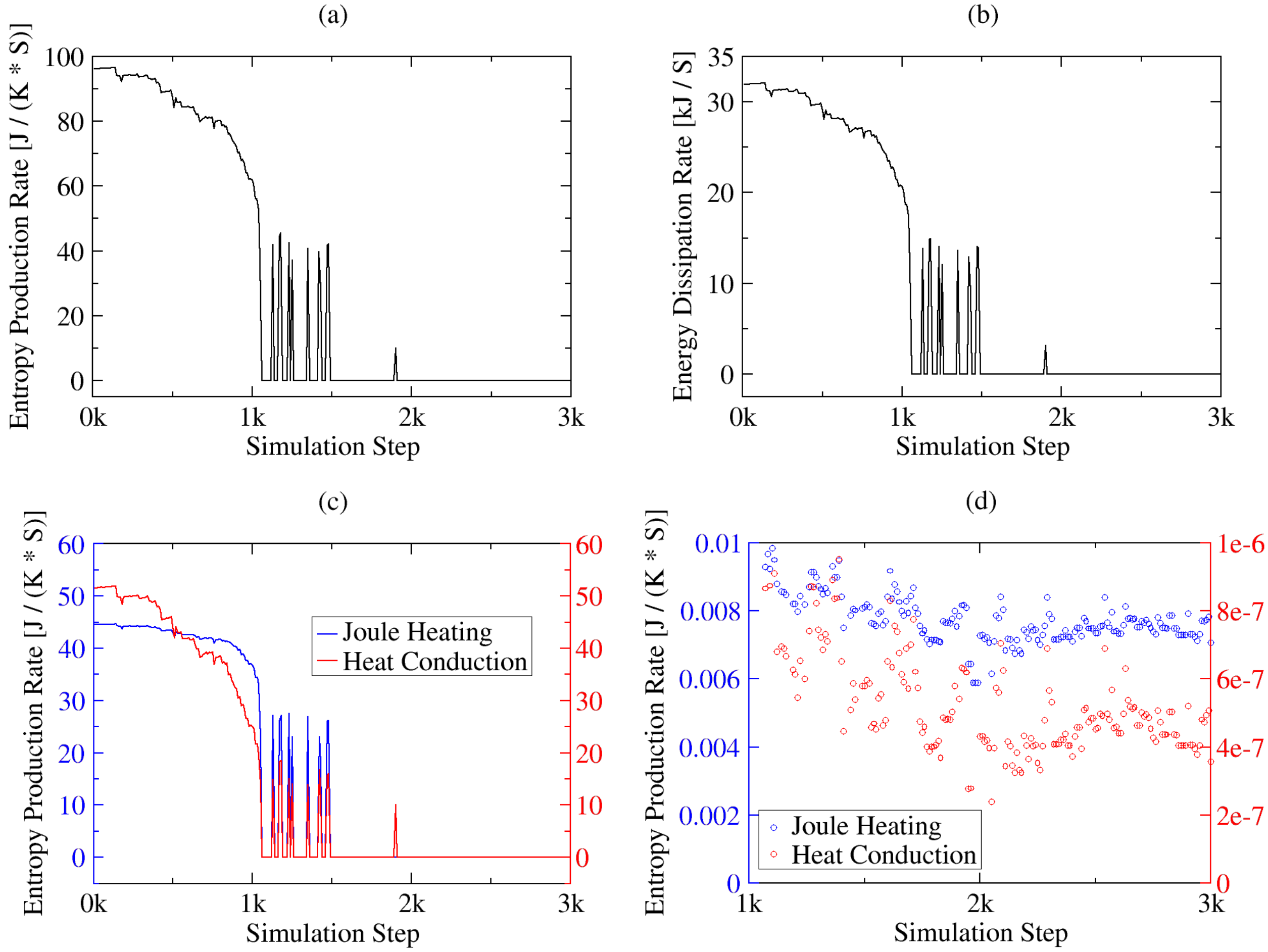

Fig. 4. Single simulation extremal behavior. (a) Entropy production rate. (b) Energy dissipation rate. (c) Entropy production rate by source. (d) Energy dissipation rate by source in the second phase with high entropy production spikes omitted.

The results of the filament morphology are shown in Figure 5 for the same simulation shown in Figure 4. From Figure 5(a), it can be seen that whenever the filament switches to the alternate mode in the second phase, the filament breaks momentarily. Across all of the simulation runs, this breakage occurs most often immediately after the transition to the second phase. The decrease in average filament width during the transition, shown between steps 800 and 1000 in Figure 5(b), reflects that the filament forms a thinner, unstable branch at the tip to quickly bridge the final gap. In the second phase, these unstable branches relax into a more stable uniform morphology, which in turn minimizes the entropy production and energy dissipation rates.

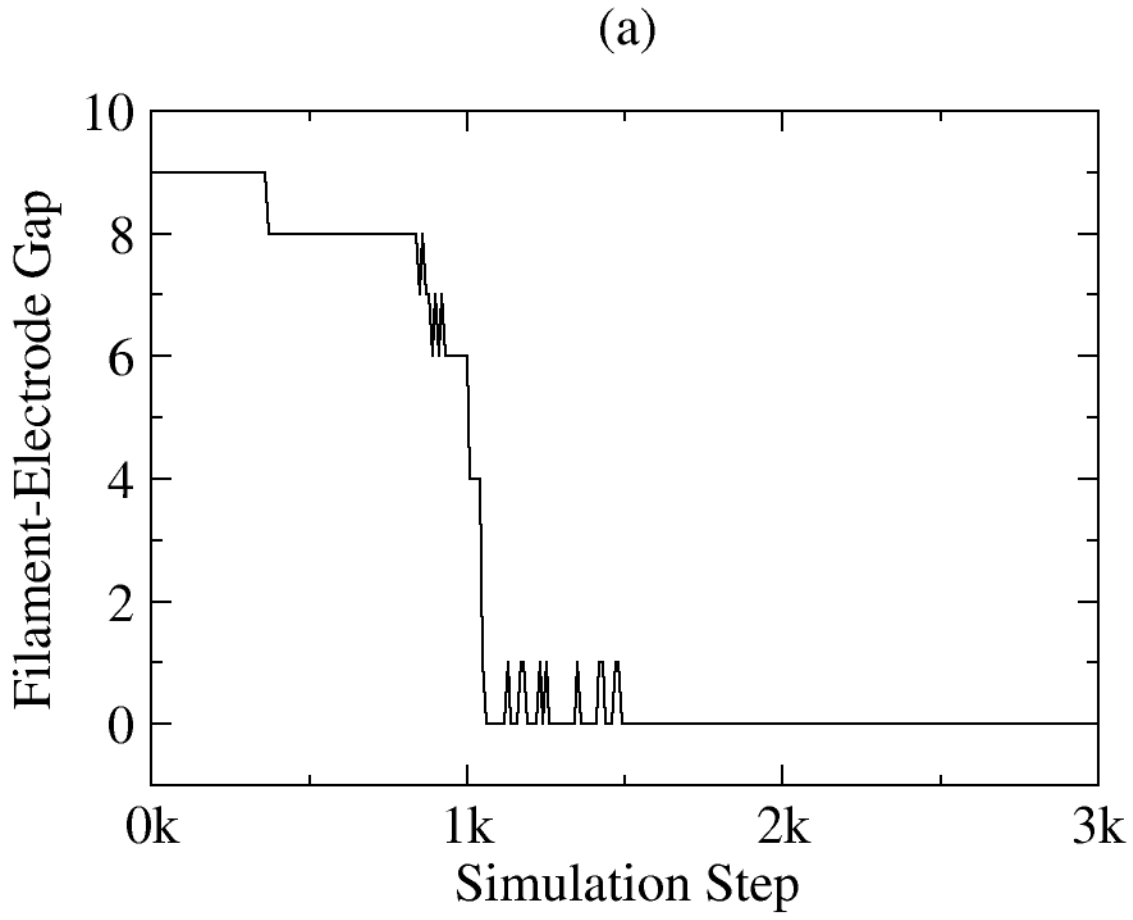

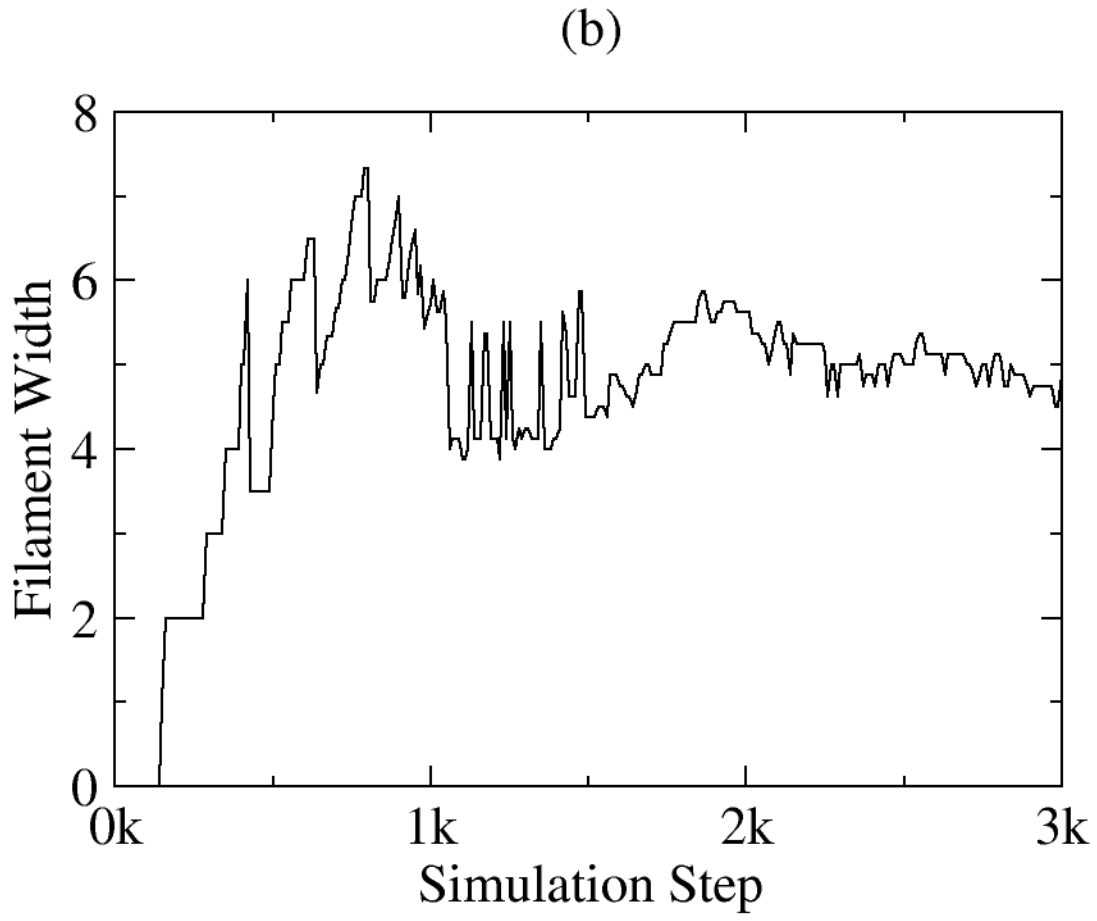

Fig. 5. Single simulation filament morphology. (a) Average distance from the filament to the active electrode, measured in units of lattice sites. (b) Average filament width in terms of number of atoms. Filament width counting does not include the seed prepatterning or the side-branches that carry negligible current.

Throughout the entire filament formation process, the system exhibits a minimization of the entropy production and energy dissipation rates. Two phases of behavior are observed. There is the initial, directed filament growth phase, where the entropy production and energy dissipation rates monotonically decrease; followed by a phase of undirected filament behavior where these rates are relatively steady. This second phase is interrupted early on by random periods of momentary filament breakage as the filament relaxes into a more uniform width morphology. In the first phase, it can be inferred from the comparable size of the Joule heating and heat conduction entropy production terms that the filament behavior is equally electrically and thermally driven, whereas in the second phase, the filament behavior is primarily electrically driven.

The initial entropy production and energy dissipation minimization are due to the decrease in device resistance as the bulk of the filament is formed, which leads to a drop in Joule heating and heat conduction. We note that while the initial state is far from the near-steady-state regime required by extremal principles, the minimizations are still consistently observed. When averaged over all simulation runs, the second phase of minimization shows undirected filament behavior. In individual runs, instability is observed immediately following the transition.

## IV. CONCLUSIONS

This study shows that the filament formation process in ECM memristors follows a principle of minimizing the entropy production and energy dissipation rates in two distinct phases. The first phase involves the initial bulk filament formation starting far from steady-state regime, and is equally electrically and thermally driven. The second phase involves the spontaneous breakage and reconnection of filament branches near the steady state, as the thin, unstable branches relax to the true steady state. Both behaviors imply that the forming process in ECM memristors can be described by extremum principles. This would, in principle, enable one to predict the steady-state filament morphology without directly simulating the growth process. Furthermore, these findings provide physical insights into the stability of the filament.

**Acknowledgments**

Computational resources were provided by The University of Memphis High-Performance Computing Center (HPCC).

**Author Contributions**

**Justin Brutger**: conceptualization (equal); investigation; data curation; formal analysis; writing - original draft preparation. **Xiao Shen**: conceptualization (equal); supervision; writing - review and editing.

**Conflict of Interest Statement**

The authors Justin Brutger and Xiao Shen have no conflicts to disclose.

**Availability of Data Statement**

The data that support the findings of this study are available from the corresponding author upon reasonable request.

**References**

[1] T. Shi et el., "A Review of Resistive Switching Devices: Performance Improvement, Characterization, and Applications," Small Struct. 2 (4), 2000109, (2021). https://doi.org/10.1002/sstr.202000109

[2] D. S. Jeong et al., “Emerging memories: resistive switching mechanisms and current status,” Rep. Prog. Phys., 75 (7), 076502, (2012). http://doi.org/10.1088/0034-4885/75/7/076502

[3] L. Chua, “Memristor-The missing circuit element,” Trans. Circuit Theory, 18 (5), 507–519, (1971). http://doi.org/10.1109/TCT.1971.1083337

[4] D. B. Strukov, G. S. Snider, D. R. Stewart, and R. S. Williams, “The missing memristor found,” Nature, 453, 80–83, (2008). http://doi.org/10.1038/nature06932

[5] E. Gale, “TiO2-based memristors and ReRAM: materials, mechanisms and models (a review),” Semicond. Sci. Technol., 29 (10), 104004, (2014). http://doi.org/10.1088/0268-1242/29/10/104004

[6] L. Gao, Q. Ren, J. Sun, S.-T. Han, and Y. Zhou, “Memristor modeling: challenges in theories, simulations, and device variability,” J. Mater. Chem. C, 9 (47), 16859-16884, (2021). https://doi.org/10.1039/D1TC04201G

[7] Y. Xiao et al., “A review of memristor: material and structure design, device performance, applications and prospects,” Sci. Technol. Adv. Mater., 24 (1), 2162323, (2023). https://doi.org/10.1080/14686996.2022.2162323

[8] E. Apollos, “Memristor Theory and Mathematical Modelling,” Int. J. Comput. Appl., 178 (27), 1–8, (2019). http://doi.org/10.5120/ijca2019919089

[9] S. Chen et al., “Electrochemical-Memristor-Based Artificial Neurons and Synapses—Fundamentals, Applications, and Challenges,” Adv. Mater., 35 (37), 2301924, (2023). https://doi.org/10.1002/adma.202301924

[10] F. M. Simanjuntak et al., “Role of nanorods insertion layer in ZnO-based electrochemical metallization memory cell,” Semicond. Sci. Technol., 32 (12), 124003, (2017). http://doi.org/10.1088/1361-6641/aa9598

[11] P. Singh, F. M. Simanjuntak, A. Kumar, and T.-Y. Tseng, “Resistive switching behavior of Ga doped ZnO-nanorods film conductive bridge random access memory,” Thin Solid Films, 660, 828–833, (2018). http://doi.org/10.1016/j.tsf.2018.03.027

[12] M. Buttberg, I. Valov, and S. Menzel, “Simulating the filament morphology in electrochemical metallization cells,” Neuromorphic Comput. Eng., 3 (2), 024010, (2023). http://doi.org/10.1088/2634-4386/acdbe5

[13] W. Wang et al., “Surface diffusion-limited lifetime of silver and copper nanofilaments in resistive switching devices,” Nat. Commun., 10 (1), 81, (2019). http://doi.org/10.1038/s41467-018-07979-0

[14] S. A. Chekol, S. Menzel, R. W. Ahmad, R. Waser, and S. Hoffmann-Eifert, “Effect of the Threshold Kinetics on the Filament Relaxation Behavior of Ag-Based Diffusive Memristors,” Adv. Funct. Mater., 32 (15), 2111242, (2021). http://doi.org/10.1002/adfm.202111242

[15] S. A. Chekol, S. Menzel, R. Waser, and S. Hoffmann-Eifert, “Strategies to Control the Relaxation Kinetics of Ag-Based Diffusive Memristors and Implications for Device Operation,” Adv. Electron. Mater., 8 (11), 2200549, (2022). http://doi.org/10.1002/aelm.202200549

[16] D. K. Kondepudi and I. Prigogine, Modern Thermodynamics: From Heat Engines to Dissipative Structures. (John Wiley & Son Ltd, New York, 1998), pp. 333-402. http://doi.org/10.1002/9781118698723

[17] J. E. Han, J. Li, C. Aron, and G. Kotliar, “Filament Dynamics in Resistive Switching,” J. Phys. Conf. Ser., 1041 (1), 012012, (2018). http://doi.org/10.1088/1742-6596/1041/1/012012

[18] V. Bertola and E. Cafaro, “A critical analysis of the minimum entropy production theorem and its application to heat and fluid flow,” Int. J. Heat Mass Transf., 51 (7), 1907–1912, (2008). http://doi.org/10.1016/j.ijheatmasstransfer.2007.06.041

[19] B. K. Ridley, “Specific Negative Resistance in Solids,” Proc. Phys. Soc., 82 (6), 954, (1963). http://doi.org/10.1088/0370-1328/82/6/315

[20] R. Waser and M. Aono, “Nanoionics-based resistive switching memories,” Nat. Mater., 6 (11), 833–840, (2007). http://doi.org/10.1038/nmat2023

[21] S. Dirkmann and T. Mussenbrock, “Resistive switching in memristive electrochemical metallization devices,” AIP Advances, 7 (6), 065006, (2017). https://doi.org/10.1063/1.4985443

[22] S. Menzel and J.-H. Hur, in Resistive Switching: From Fundamentals of Nanoionic Redox Processes to Memristive Device Applications, edited by D. Ielmini and R. Waser (John Wiley & Sons, Hoboken, NJ 2016), pp. 395–436. http://doi.org/10.1002/9783527680870.ch14

[23] S. Dirkmann et al., “Kinetic simulation of filament growth dynamics in memristive electrochemical metallization devices,” J. Appl. Phys., 118 (21), 214501, (2015). https://doi.org/10.1063/1.4936107

[24] B. I. Kharisov, O. V. Kharissova, and U. Ortiz-Mendez, CRC Concise Encyclopedia of Nanotechnology, CRC Press, p. 726, 2016.

[25] P. Enghag, Encyclopedia of the Elements: Technical Data - History - Processing - Applications, Wiley, p. 125, 2004.

[26] M. Saeedian, M. Mahjour-Shafiei, E. Shojaee, and M. R. Mohammadizadeh, “Specific heat capacity of TiO2 nanoparticles,” arXiv:1307.7555, 2013.

[27] S. I. Abu-Eishah, Y. Haddad, A. Solieman, and A. Bajbouj, “A new correlation for the specific heat of metals, metal oxides and metal fluorides as a function of temperature,” Lat. Am. Appl. Res. 113 (4), 257-264 (2004).